\documentclass{easychair}

\usepackage{listings}
\usepackage{booktabs}
\usepackage{soul,color}
\usepackage{subcaption}
\usepackage{orcidlink}

\title{From Tweet to Theft:\\ Tracing the Flow of Stolen Cryptocurrency}

\authorrunning{Cola, Mazza, and Tesconi}

\titlerunning{From Tweet to Theft}

\author{
    Guglielmo Cola\inst{1}\orcidlink{0000-0003-2890-723X}
\and
    Michele Mazza\inst{1}\orcidlink{0000-0003-1874-3753}
\and
    Maurizio Tesconi\inst{1}\orcidlink{0000-0001-8228-7807}
}

\institute{
  Institute of Informatics and Telematics (IIT)\\ National Research Council (CNR), Via G. Moruzzi 1, 56124, Pisa, Italy\\ 
  \email{\{g.cola, m.mazza, m.tesconi\}@iit.cnr.it}
}

\begin{document}

\maketitle

\noindent\textbf{\hl{Note to Readers}:} This is a revised version of the paper originally presented at the ITASEC 2023 conference and included in CEUR Workshop Proceedings\footnote{\url{https://ceur-ws.org/Vol-3488/paper11.pdf}}. The revision includes new information disclosed by Arkham Intelligence, a leading blockchain analytics firm. Specifically, it has been revealed that the address we referred to as ``forwarding address'' is actually part of the cryptocurrency swap service SimpleSwap. This information enhances our understanding of the scam's operations and the methods used to obfuscate the trail of stolen cryptocurrency.

\begin{abstract}
This paper presents a case study of a cryptocurrency scam that utilized coordinated and inauthentic behavior on Twitter. In 2020, 143 accounts sold by an underground merchant were used to orchestrate a fake giveaway. Tweets pointing to a fake blog post lured victims into sending Uniswap tokens (UNI) to designated addresses on the Ethereum blockchain, with the false promise of receiving more tokens in return. Using one of the scammer's addresses and leveraging the transparency and immutability of the Ethereum blockchain, we traced the flow of stolen funds through various addresses, revealing the tactics adopted to obfuscate traceability. The final destination of the funds involved two deposit addresses. The first, managed by a well-known cryptocurrency exchange, was likely associated with the scammer's own account on that platform and saw deposits exceeding \$3.5 million. The second address was linked to a popular cryptocurrency swap service. These findings highlight the critical need for more stringent measures to verify the source of funds and prevent illicit activities.
\end{abstract}


\section{Introduction}
\label{sect:introduction}

The growing adoption of cryptocurrencies has put under the spotlight the risks for the users of such blockchain-based technologies. Due to the decentralized and immutable nature of distributed ledgers, users must exercise a high degree of caution to protect themselves from potential fraud. Twitter, with its large user base, has become a popular target for scammers seeking to reach inexperienced cryptocurrency users. One such scam is the advance-fee scam, where the victims are lured into sending funds to the scammer with the fake promise of receiving a larger sum in return. Fake giveaways, a form of advance-fee scam, are particularly prevalent on social media. They are generally based on false content designed to convince users that a well-known individual is giving away money, for example to celebrate a specific event or milestone. Coordinated and inauthentic behavior from fake accounts, commonly known as CIB, enable scammers to spread their deceptive invitation to a large audience, making it easier for them to reach vulnerable and unsuspecting users.

The investigation presented in this paper stems from a previous work related to CIBs produced by fake accounts~\cite{mazza2022ready}. In that work, we proposed a novel approach to spot fake accounts for sale and then monitor their activity. As a result, over five million tweets and four coordinated campaigns were found throughout 2020. One of these CIBs was aimed at promoting a fake giveaway related to the Uniswap token (UNI). A group of fake accounts advertised a link to a fraudulent blog post, which in turn described the fake giveaway and lured users into sending their tokens to a designated address on the Ethereum blockchain. These URLs are no longer accessible at present, still an example of their content can be retrieved by using the Internet Archive. Also, during our research in~\cite{mazza2022ready}, we were able to capture a screenshot and identify one of the Ethereum addresses used by the scammer. 

In this paper, we leverage the immutable and public nature of the Ethereum blockchain to investigate the flow of ill-gotten UNI tokens, starting from that single address linked to the scammer. Our analysis reveals some of the strategies used by scammers to capture funds and transfer them to cryptocurrency exchanges, where the stolen funds cannot be traced anymore by using plain blockchain analysis. The presented findings can foster further research in the area, which can possibly lead to the development of automated techniques for quickly identifying malicious addresses and mitigating the harm to cryptocurrency and social media users.

The paper is organized as follows. In the following Section we briefly present some of the most relevant studies in the field of cryptocurrency scams, with a focus on the advance-fee scam form. In Section~\ref{sec:twitter} we provide some details on the CIB that lured victims into sending UNI tokens to the scammer. Next, in Section~\ref{sec:blockchain} we show the blockchain-based analysis of the flow of UNI tokens, which uncovers the technique used by the scammer to deposit the stolen funds to a well-known cryptocurrency exchange and a popular cryptocurrency swap service.
Section~\ref{sec:conclusions} concludes the paper and suggests avenues for future research.

\section{Related work}

The last few years have witnessed increased interest in blockchain-based technologies, not only from the general public but also from institutional investors~\cite{HUANG2022110856}. This rise in popularity has led to a corresponding increase in the risks associated with cryptocurrencies, as various types of scams have been exploited by malicious actors~\cite{bartoletti2021cryptocurrency}. Examples of scams that have been discussed by the literature include 
advance-fee scams~\cite{phillips20}, fake exchange apps~\cite{xia2020characterizing}, fake initial coin offerings (ICOs)~\cite{zetzsche2019ico}, phishing attacks~\cite{holub18}, Ponzi schemes~\cite{vasek2019analyzing, bartoletti2020dissecting}, and smart contract honeypots~\cite{ferreira19}.

Cybercriminals have exploited the ``pseudonymity'' feature of blockchains, which allows them to conceal their identities while carrying out illicit activities. However, the transparency and immutability of public blockchains also enable the development of techniques to trace the flows produced by such illicit activities. Various techniques have been proposed for this purpose, including graph analysis and machine learning~\cite{wu2020phishers, farrugia2020detection}, with promising results.

Social media and the use of fake accounts have greatly facilitated the spread of misleading contents aimed at targeting unsuspecting cryptocurrency users~\cite{mazza2022ready}.
One particular scam that has gained attention is the advance-fee scam, where the victims are lured into sending an amount of money as a fee with the false promise of receiving a greater return~\cite{bartoletti2021cryptocurrency}. In the field of blockchain-based analysis of advance-fee scams, a relevant contribution was made by~\cite{phillips20}. Using DBSCAN clustering on the content of scam websites, they found out that the same entities are running multiple instances of similar scams, as revealed also by their blockchain activity. These bad actors can even fabricate ad-hoc blockchain activity to pretend their promises are genuine. The authors also reported that cryptocurrency exchanges were the most common destination for funds obtained through such scams, followed by gambling platforms.

While illicit transactions should not overshadow the potential benefits that blockchain technologies can offer to various application areas~\cite{monrat2019survey}, the associated risks cannot be ignored. Our work can contribute to a deeper understanding of advance-fee scams in the context of cryptocurrency, and ultimately help devise proper measures to mitigate their impact on users.

\section{Coordinated behavior from fake Twitter accounts}
\label{sec:twitter}

\begin{figure}[t]
  \centering
  \includegraphics[width=1.0\linewidth]{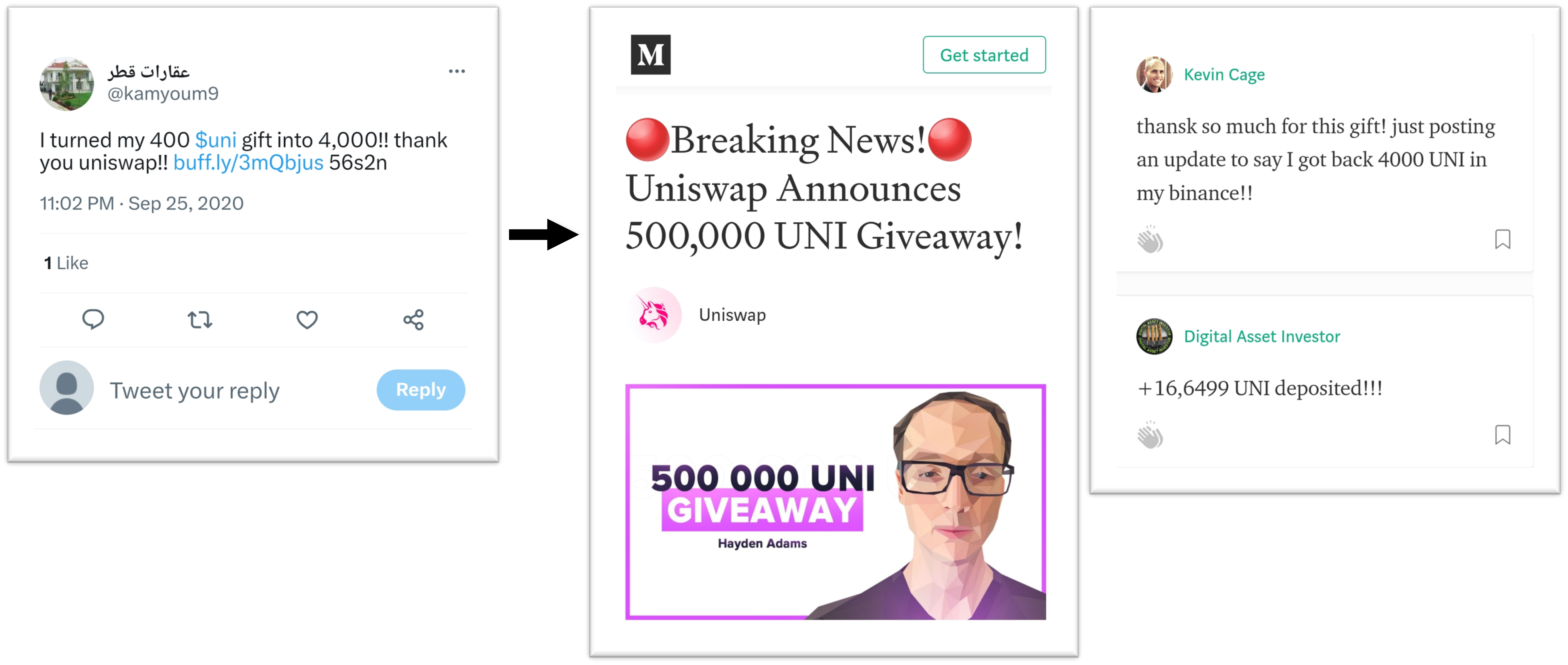}
  \caption{Tweet from a fake account and fake giveaway blog post}
  \label{fig:fake_blog}
\end{figure}

\begin{table}[t]
    \centering
    \caption{Most used hashtags in tweets advertising the fake giveaway}
    \begin{tabular}{lcc}
        \toprule
        Hashtag & Tweets & Accounts\\
        \midrule
        \#uniswap & 12,169 & 101 \\
        \#defi & 9,866 & 78\\
        \#uni & 1,197 & 14\\
        \#sushiswap & 623 & 7\\
        \#yfi & 524 & 12\\
        \#crypto & 466 & 38\\
        \#DeFi & 454 & 14\\
        \#bitcoin & 278 & 33\\
        \#relax & 212 & 2\\
        \#Crypto & 187 & 30\\
        \bottomrule
    \end{tabular}
    \label{tab:hashtags}
\end{table}

The analysis presented in the paper stems from a broader study on \textit{coordinated inauthentic behavior} (CIB) that we described in~\cite{mazza2022ready}.  Fake accounts, i.e., accounts that hide the real identities of the people running them~\cite{mazza2022investigating}, were identified on website buyaccs.com and then monitored over the course of one year to detect coordinated and inauthentic behavior (CIB). 

One of the four CIBs that emerged consisted in the 143 accounts that orchestrated the Uniswap-related fake giveaway. 
These accounts were virtually inactive throughout 2020, except for the second part of September, during which they shared 146,546 tweets.
The activity turned out to be a scam operation which exploited the launch of the UNI token, occurred a few days before these tweets started. To reach potential victims, the fake accounts used both hashtags strictly related to the UNI token and more generic hashtags related to the decentralized finance paradigm and other cryptocurrencies, as shown in Table~\ref{tab:hashtags}. The strategy behind this scam operation features tweets as a starting point. In fact, in their tweets the fake accounts claimed to have multiplied by ten times their amount of UNI tokens. Moreover, the tweets featured a URL (often shortened through the buffer.com service) pointing to articles that were visually identical to an article posted on medium.com. An example of a fake tweet and part of the blog post is shown in Figure~\ref{fig:fake_blog}.
The article was about a UNI token giveaway and included a second URL to reach the giveaway website, which invited users to send their UNI tokens to a designated address on the Ethereum blockchain. Furthermore, instructions were given on how to multiply the tokens: for every token sent to the address on the website, one would receive back ten times as many. As shown in Figure~\ref{fig:fake_blog}, there was also a comment section with several fake positive feedback. Thus, victims of the scam were tricked into sending their UNI tokens to the address, with the false promise of receiving more tokens in return.
Although the strategy sounds quite simple, relying on timing and prepared content to exploit the hype around a specific event, it has also been proven effective, as shown in the following Section. 

Twitter systems seemed to be ineffective in countering the operation on its own platform, as despite the large volume of tweets produced in a limited time window, 97 accounts out of 143 were still active at the end of 2020. Moreover, 48 accounts result active as of February 2023, and their tweets related to the scam have not been removed. The links to the fake post are broken, however its content can be retrieved on the Internet Archive.

\section{Blockchain analysis}
\label{sec:blockchain}

Ethereum is an open-source and decentralized blockchain network that was designed to enable the creation of smart contracts, which in turn can be used to build decentralized applications (DApps). Ethereum's native cryptocurrency is ether (ETH), which is used to pay for ``gas fees'', namely the fees associated with executing transactions and smart contracts. One key feature of Ethereum is the ability to ``mint'' and transfer fungible tokens, which could be used to represent new assets or a specific utility according to the ERC-20 standard. Non-fungible tokens, known as NFTs, can be minted as well, according to the ERC-721 standard. Each address on the Ethereum blockchain must contain at least a small amount of ether to interact actively with the blockchain (i.e., send ETH/tokens to other addresses or request execution to a smart contract). Hence, each address may contain ETH and/or tokens.

The UNI token is an example of an ERC-20 fungible token. More specifically, it was designed as a \textit{governance token}: possessors of UNI gain the ability to vote for decisions regarding Uniswap. The latter was the first popular decentralized exchange, i.e., a DApp that allows users to swap tokens while retaining ownership of funds and without the need for trusting a centralized platform like Coinbase or Binance. Starting from September 2020, the addresses that interacted with Uniswap before September 1, 2020, were eligible to claim a specific amount of UNI tokens. The scammer exploited this event to organize the fake giveaway presented in the previous Section.

In the following we show how we traced the stolen funds starting from a single address belonging to the scammer and using blockchain analysis.

\subsection{Overview of address movements}

\begin{figure}[t]
  \centering
  \includegraphics[width=0.8\linewidth]{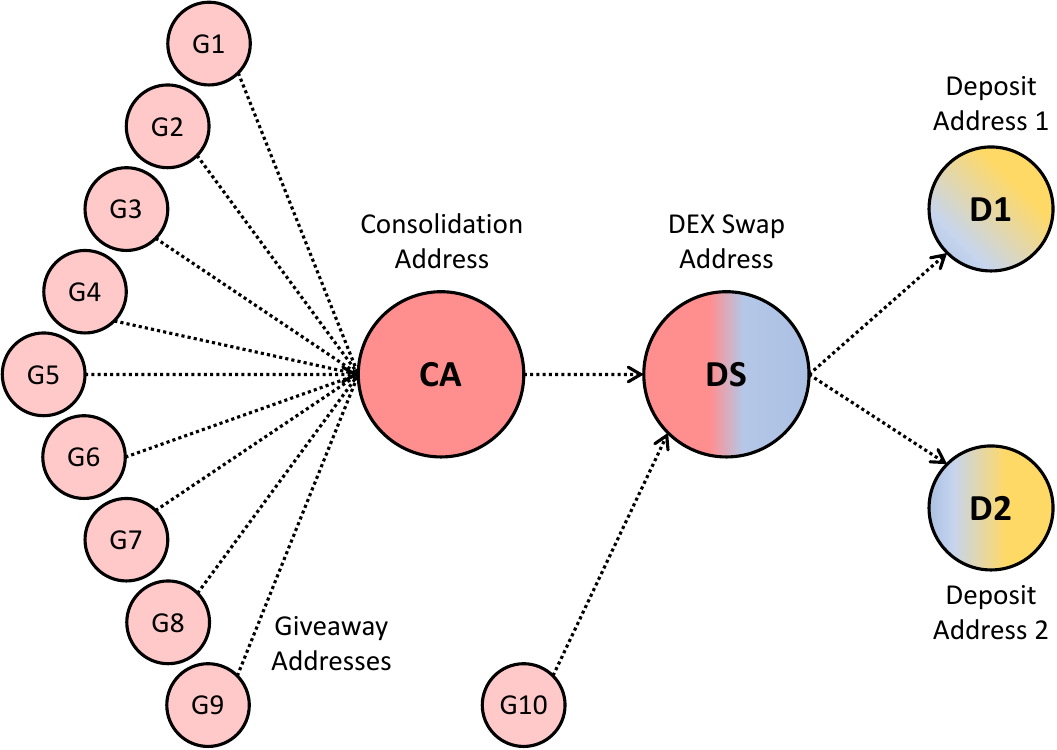}
  \caption{Overview of the addresses used in the scam}
  \label{fig:flow_scheme}
\end{figure}

Our blockchain analysis started from a single address used by the scammer to receive the ``advance-fee'' in UNI tokens from victims. Using the popular block explorer Etherscan and its APIs, we identified the amount of UNI received by the address and the destination of such tokens. Following the flow of UNI tokens, we uncovered the set of addresses used by the scammer, as shown by the graph in Figure~\ref{fig:flow_scheme}.

The graph displays the flow of UNI tokens, with single addresses as nodes and edges representing transactions to another address. The initial address from which our analysis started is labeled as G5: it belongs to a group of ``giveaway addresses'' that were advertised directly to victims so as to receive UNI tokens. It can be seen that ten addresses were used in this way: nine of them (G1-G9) sent their tokens to another address that we named ``Consolidation address'' (CA). CA then sent the tokens to a new address, which we named ``Dex Swap Address'' (DS). DS also received funds directly from another giveaway address, G10, before converting all the UNI into ETH by using a decentralized exchange. After, it sent approximately two-thirds of the ETH to an exchange deposit address (D1) and the remaining to a cryptocurrency swap service deposit address (D2).

From a temporal point of view, most of the UNI tokens were received on the giveaway addresses between September 18 and September 28, 2020. During these days, the giveaway addresses were advertised on Twitter and thus received UNI tokens from victims. On September 28, funds from all the giveaway addresses except G10 were moved to CA. All the remaining transactions occurred on November 3, 2020, when the scammer amassed funds from CA and G10 to DS, before swapping them for ETH and making two separate deposits to centralized services (D1 and D2).

Further details and clarifications on each step are provided in the following subsections.

\subsection{Giveaway addresses}

\begin{figure}[t]
  \centering
  \includegraphics[width=0.9\linewidth]{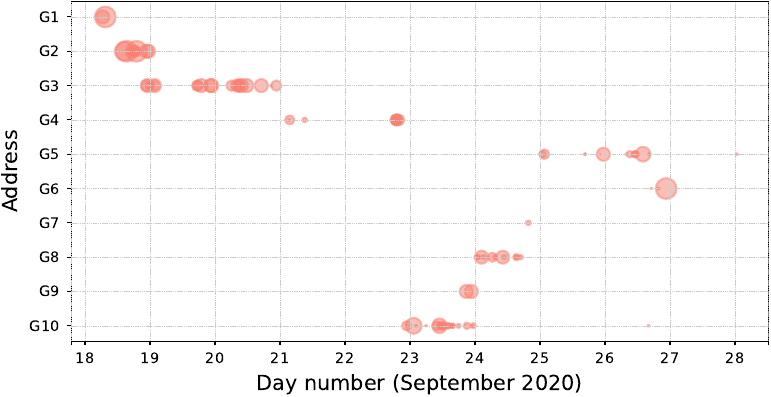}
  \caption{Temporal distribution of UNI token deposits to giveaway addresses}
  \label{fig:ga_txs}
\end{figure}

\begin{table}[t]
    \centering
    \caption{Outbound UNI token transactions from giveaway addresses}
    \begin{tabular}{lccr}
        \toprule
        Date (UTC) & From & To & Amount\\
        \midrule
2020-09-28 07:51	&	G1	&	CA	&	1,399.9	\\
2020-09-28 07:53	&	G2	&	CA	&	4,581.2	\\
2020-09-28 07:55	&	G3	&	CA	&	5,865.1	\\
2020-09-28 07:58	&	G4	&	CA	&	1,651.1	\\
2020-09-28 07:59	&	G5	&	CA	&	1,567.5	\\
2020-09-28 08:02	&	G6	&	CA	&	1,017.9	\\
2020-09-28 08:04	&	G7	&	CA	&	52.4	\\
2020-09-28 08:05	&	G8	&	CA	&	1,430.8	\\
2020-09-28 08:10	&	G9	&	CA	&	800.0	\\
2020-11-03 13:12	&	G10	&	DS	&	2,443.4	\\
        \bottomrule
    \end{tabular}
    \label{tab:txGA}
\end{table}
 
Most of the giveaway addresses were labeled on Etherscan as ``Phish/Hack'', meaning that users reported them as malicious. Among G1-G10, only G4 and G6 did not receive such label, however, their activity pattern strongly suggest that they were part of the addresses used to collect funds from victims. For each of the giveaway addresses, the pattern was the same: receiving UNI tokens from other addresses before sending all the tokens to another address (CA for G1-G9, DS for G10). 

More detail on the timing of the deposits to giveaway addresses is revealed in Figure~\ref{fig:ga_txs}. In the plot, each dot represents a UNI token deposit to a specific address, with size proportional to the amount of tokens. It can be observed that there is almost no temporal overlap between the deposits to different addresses. 
This suggests that the scammer might have changed the address advertised in the blog post so as to elude user reports of malicious activity, which might have informed potential victims. 

As shown in Figure~\ref{fig:flow_scheme}, G1-G9 sent their tokens to CA, whereas G10 sent its UNI tokens directly to DS. More information on outbound transactions from giveaway addresses is provided in Table~\ref{tab:txGA}. The deposits from G1-G9 to CA occurred in a very short timeframe, about 20 minutes in the morning of September 28. Instead, funds from G10 were moved to DS on November 3. 

On September 28, the giveaway addresses contained a total 20,809 UNI tokens. If we consider the UNI value of \$4.78 provided by crypto data aggregator CoinGecko (\url{www.coingecko.com}) for September 28, it turns out that the scammer managed to collect around \$100,000 worth of UNI in just ten days.

\subsection{Path to deposit addresses}

\begin{table}[t]
    \centering
    \caption{Remaining transactions to move the funds to the deposit addresses (D1 and D2)}
    \begin{tabular}{lcccr}
        \toprule
        Date (UTC) & From & To & Crypto & Amount\\
        \midrule
2020-11-03 13:18	&	CA	&	DS	&	UNI	&	18,365.8	\\
\midrule
2020-11-03 15:24	&	DS	&	1inch	&	UNI	&	20,809.3	\\
2020-11-03 15:24	&	1inch	&	DS	&	ETH	&	115.8	\\
\midrule
2020-11-03 15:26	&	DS	&	D1	&	ETH	&	40.0	\\
2020-11-03 15:28	&	DS	&	D2	&	ETH	&	40.0	\\
2020-11-03 15:43	&	DS	&	D1	&	ETH	&	48.2	\\
        \bottomrule
    \end{tabular}
    \label{tab:txCADS}
\end{table}

The list of transactions involving CA and DS is shown in Table~\ref{tab:txCADS}. As mentioned in the previous subsection, CA received 18,365.8 UNI from G1-G9. These tokens remained dormant until November 3, 2020, when they were transferred to DS. As shown in Table~\ref{tab:txGA}, on the same day (just six minutes earlier) DS also received 2,443.4 UNI tokens from G10. A couple of hours later, the UNI tokens in DS were swapped for around 115.8 ETH using the popular DEX aggregator named 1inch. A DEX aggregator connects automatically to multiple decentralized exchanges, in order to find the best ``route'' for a given trade. As mentioned before, decentralized exchanges, like Uniswap, allow users to trade cryptocurrencies without going through centralized exchanges, as the trade is managed by a smart contract executed on the blockchain. An additional 12.4 ETH were obtained in DS by swapping two different ERC-20 tokens, sushiToken (SUSHI) and yearn.finance (YFI), whose origin is beyond the scope of this investigation. The total 128.2 ETH were then split: 88.2 ETH were sent to a first deposit address (D1) with two transactions of 40.0 and 48.2 ETH, while the remaining 40.0 ETH were sent to another deposit address (D2).

\subsection{D1 -- Exchange deposit address}

\begin{table}[htbp]
    \centering
    \caption{Value of deposits in D1}
    \begin{tabular}{clr}
      \toprule
    \multicolumn{2}{c}{Top 5 Tokens} & Value \$ \\
      \midrule
1	&	USDC	&	1,474,369	\\
2	&	ETH	&	1,169,189	\\
3	&	USDT	&	380,879	\\
4	&	REN	&	185,863	\\
5	&	TUSD	&	122,921	\\
        \midrule
        \multicolumn{2}{l}{Total D1} & 3,564,959\\
      \bottomrule
    \end{tabular}
    
    \label{tab:D1}
\end{table}

\begin{figure}[t]
  \centering
  \includegraphics[width=0.9\linewidth]{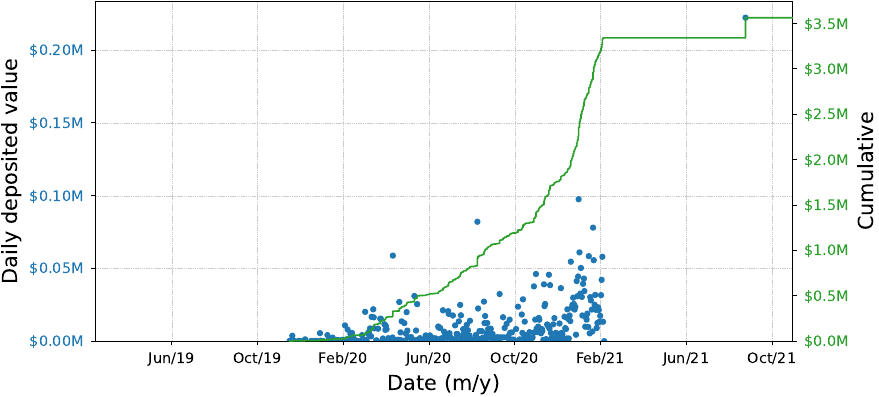}
  \caption{Daily and cumulative deposits made to D1}
  \label{fig:D1_deposits}
\end{figure}

The scammer(s) deposited 88.2 ETH to address D1.
The transactions involving this address reveal patterns that are typical of an exchange deposit address. A centralized cryptocurrency exchange, also known as CEX, is a platform that acts as an intermediary between buyers and sellers, and that makes money mainly through commissions and fees. Popular CEXes include Binance, Coinbase, and Crypto.com. To start using a CEX, users have to deposit fiat currrency or cryptocurrency, which can then be exchanged for other fiat/crypto. For crypto deposits, CEXes provide users with a dedicated deposit address on the blockchain. For instance, if the user asks to deposit ETH or ERC-20 tokens, the CEX provides an address on the Ethereum blockchain where the user can deposit their ETH/tokens. After the deposit is made, the CEX credits the deposited amount to the user's balance on the platform, which can then be exchanged for the other fiat currencies or crypto tokens available on the CEX. 

The deposit address is generated by the CEX, which controls the keys needed to move funds out of that address. Therefore, when a user deposits crypto to such an address, the effective ownership of the cryptocurrency is transferred to the CEX, while the user becomes a creditor. The cryptocurrency received in a deposit address is typically moved by the CEX to another address, where the cryptocurrencies received from multiple users are aggregated for easier management. More information on how deposit addresses are managed is provided in this blog post\footnote{\url{www.binance.com/en/blog/from-cz/transparency-on-wallets-at-binance-4080204794657829130}} from popular CEX Binance.

Address D1 received over 1,500 deposits.
Most of the cryptocurrency deposited was later transferred to Ethereum addresses that are publicly labeled as belonging to Binance. Also, the addresses received small amounts of ETH to be used for gas fees from a smart contract created by Binance itself. These patterns strongly suggest that D1 is controlled by Binance and allowed the scammer(s) to deposit their ETH and ERC-20 tokens on that popular CEX. By using a CEX, the scammer was able to break the traceability of funds on the blockchain. Presumably, funds were later withdrawn to a different address on the blockchain or traded for fiat currency (e.g., US dollars or euros) and then sent to a bank account. Indeed, differently from the transactions happening on the blockchain, what happens within a CEX cannot be studied publicly, as funds from all users are merged into a limited set of adddresses belonging to the CEX itself, and there is no way to link the initial deposit address to a withdrawal address.

To evaluate the scale of the scammer's activity, Table~\ref{tab:D1} presents the top five tokens deposited to D1 ranked by their aggregated USD value. The USD value of each deposit was calculated using the exchange rate at the time of deposit. The total USD-equivalent value deposited is also shown, considering ETH as well as all the ERC-20 tokens deposited to D1.
The total USD value is remarkable, amounting to over \$3.5 million. 
Top used tokens show a prevalence of stablecoins (USDC, USDT, TUSD), i.e., tokens that aim to be pegged to the value of the US dollar and thus protect users from the typical volatility of the crypto market.
All the tokens shown are highly popular in the context of decentralized finance.

A temporal view of the scammer's activity related to D1 is provided in Figure~\ref{fig:D1_deposits}. Each blue dot shows the aggregated value of the deposits made on a single day (value on the left y axis). Instead, the green line shows the cumulative USD-equivalent value of deposits over time (value on the right y axis). Deposits were aggregated daily for better visualization. 
D1 was mostly active between November 2019 and February 2021, with the exception of a very large USDT transaction (worth around \$220,000) on August 26, 2021. 
Daily deposits had an average value of about \$10,450.
It should be noted that the distribution of daily deposited values is highly skewed, as there are exceptionally high transactions on given days as well as a large number of relatively small deposits. 

\begin{figure}[t]
  \centering
  \includegraphics[width=0.9\linewidth]{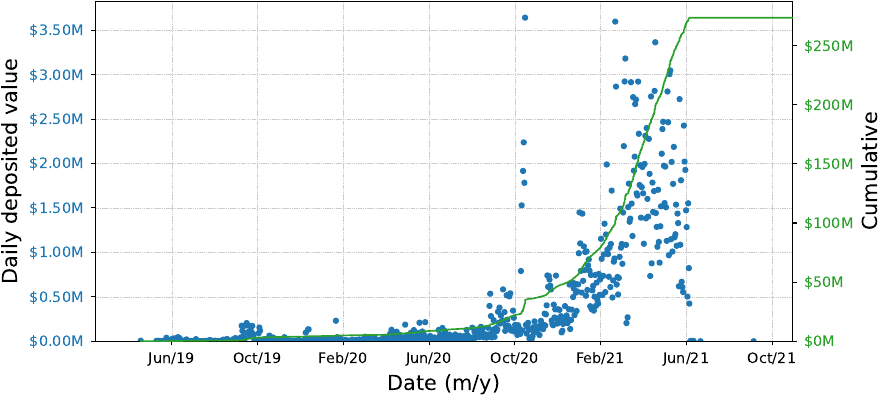}
  \caption{Daily and cumulative deposits made to SimpleSwap's deposit address on Binance}
  \label{fig:SS_deposits}
\end{figure}

\subsection{D2 -- Swap service deposit address}

Address D2 received 40 ETH, which were entirely transferred to another address within just 32 minutes. After these two events, D2 became completely inactive. The address that received the funds is another deposit address owned by Binance, thus indicating it is linked to a Binance customer. Initially, we believed address D2 was used as a forwarding address to slightly obfuscate the trail of funds. However, the blockchain analytics firm Arkham Intelligence\footnote{\url{https://platform.arkhamintelligence.com/}} disclosed that D2 is a deposit address of the cryptocurrency swap service SimpleSwap. SimpleSwap allows users to swap a wide variety of cryptocurrencies without the need to create an account. Users send their cryptocurrency to a specifically designated address; shortly thereafter, the cryptocurrency resulting from the swap is sent to a user-specified address, which could be on Ethereum or another blockchain. The funds sent to a deposit address (D2 in this study) are quickly aggregated by SimpleSwap into a deposit address controlled by a CEX, such as Binance, to facilitate future swap operations. SimpleSwap thus uses Binance and other popular CEXes to hold custody of their funds.

Figure~\ref{fig:SS_deposits} shows the volume of SimpleSwap's deposit address on Binance. There are over 79,000 deposits, presumably from multiple legitimate users. The total deposited value in USD terms amounts to over \$270 million. As all funds were merged, it is practically impossible to understand which of these funds came from the scammer analyzed in this study. Furthermore, it is impossible to determine which token (and on which blockchain) the original 40 ETH sent by the scammer were converted into.

\section{Conclusions and future work}
\label{sec:conclusions}

In this case study, we presented a blockchain-based investigation into a fake giveaway of Uniswap (UNI) tokens that was promoted through coordinated and inauthentic behavior on social media. As we have demonstrated, the scammer lured inexperienced cryptocurrency users into sending around \$100,000 worth of UNI to a set of designated addresses on the Ethereum blockchain between September 18 and September 28, 2020. On November 3, 2020, the UNI tokens were aggregated into a single ``Consolidation'' address before being sent to a new address, to which we have assigned the pseudonym ``DEX Swap''. This address was used to swap the UNI tokens for ETH via a decentralized finance platform. Finally, the resulting ETH was split into two separate addresses (D1 and D2), which we later identified as deposit addresses controlled by the popular exchange Binance and cryptocurrency swap service SimpleSwap, respectively. This enabled the scammers to obfuscate the trail of stolen cryptocurrency, and they were presumably able to cash out the funds or move them to different addresses on Ethereum or even different blockchains.

This study shows how the transparency of public blockchains can facilitate the linkage of illicit activities starting from a single address labeled as malicious. The UNI scam proved to be just the tip of the iceberg, as the aggregate activity on D1 revealed a volume of deposits worth over \$3.5 million. 
Even though we cannot infer that all of these funds were obtained illicitly and that the scammer managed to withdraw their funds from Binance, the potential impact revealed by our analysis is concerning. In general, it could be difficult for exchanges to determine with certainty whether deposited funds were obtained illicitly, especially if the scammer adopts sophisticated techniques to obfuscate the origin of their funds. Nevertheless,
these findings emphasize the need for centralized exchanges to adopt more stringent measures to prevent illicit activities and protect their users. This is especially critical for deposits originating from cryptocurrency swap services, as these platforms can obscure fund origins due to their no-account trading feature.
Automated blockchain-based analysis could be a key tool to promptly identify suspicious activity.

In future work, we plan to extend the graph of transactions related to this and other scams in order to identify common and detectable patterns. Also, it would be interesting to explore automated techniques to detect suspicious deposit addresses, which could help centralized exchanges in complying with stricter policies.

\section*{Acknowledgments}
  This work was partially supported by: the European Union – Horizon 2020 Program under the scheme “INFRAIA-01-2018-2019 – Integrating Activities for Advanced Communities”, Grant Agreement n.871042,  “SoBigData++: European Integrated Infrastructure for Social Mining and Big Data Analytics” (\url{http://www.sobigdata.eu}); project SERICS (PE00000014) under the NRRP MUR program funded by the EU - NGEU; Regione Toscana, which funded this project (Project INTERROGATE) within the framework of the “Por Fse 2014-2020 Investimenti a favore della crescita, dell'occupazione e del futuro dei giovani Fondo Sociale Europeo”. Decreto di finanziamento DD 21607 del 29/11/2021.

\begin{figure}[h]
  \centering
  \includegraphics[width=0.9\linewidth]{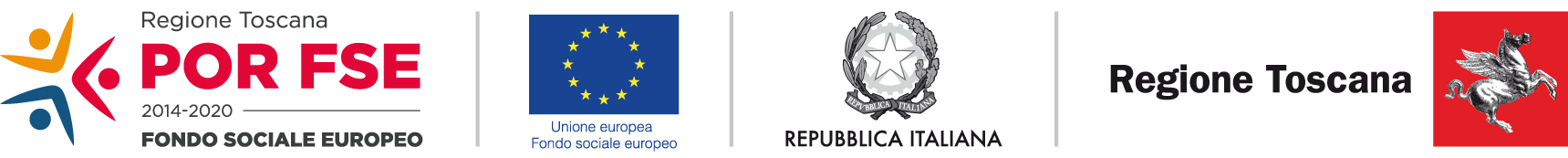}
  \label{fig:logo_tos}
\end{figure}




\bibliographystyle{plain}
\bibliography{references}

\begin{thebibliography}{10}

\bibitem{bartoletti2020dissecting}
Massimo Bartoletti, Salvatore Carta, Tiziana Cimoli, and Roberto Saia.
\newblock Dissecting ponzi schemes on ethereum: Identification, analysis, and impact.
\newblock {\em Future Generation Computer Systems}, 102:259--277, 2020.

\bibitem{bartoletti2021cryptocurrency}
Massimo Bartoletti, Stefano Lande, Andrea Loddo, Livio Pompianu, and Sergio Serusi.
\newblock Cryptocurrency scams: analysis and perspectives.
\newblock {\em IEEE Access}, 9:148353--148373, 2021.

\bibitem{farrugia2020detection}
Steven Farrugia, Joshua Ellul, and George Azzopardi.
\newblock Detection of illicit accounts over the ethereum blockchain.
\newblock {\em Expert Systems with Applications}, 150:113318, 2020.

\bibitem{holub18}
Artsiom Holub and Jeremiah O'Connor.
\newblock {COINHOARDER}: Tracking a ukrainian bitcoin phishing ring {DNS} style.
\newblock In {\em 2018 APWG Symposium on Electronic Crime Research (eCrime)}, pages 1--5. IEEE, 2018.

\bibitem{HUANG2022110856}
Xiaoran Huang, Juan Lin, and Peng Wang.
\newblock Are institutional investors marching into the crypto market?
\newblock {\em Economics Letters}, 220:110856, 2022.

\bibitem{mazza2022investigating}
Michele Mazza, Marco Avvenuti, Stefano Cresci, and Maurizio Tesconi.
\newblock Investigating the difference between trolls, social bots, and humans on twitter.
\newblock {\em Computer Communications}, 196:23--36, 2022.

\bibitem{mazza2022ready}
Michele Mazza, Guglielmo Cola, and Maurizio Tesconi.
\newblock Ready-to-(ab)use: From fake account trafficking to coordinated inauthentic behavior on twitter.
\newblock {\em Online Social Networks and Media}, 31:100224, 2022.

\bibitem{monrat2019survey}
Ahmed~Afif Monrat, Olov Schel{\'e}n, and Karl Andersson.
\newblock A survey of blockchain from the perspectives of applications, challenges, and opportunities.
\newblock {\em IEEE Access}, 7:117134--117151, 2019.

\bibitem{phillips20}
Ross Phillips and Heidi Wilder.
\newblock Tracing cryptocurrency scams: Clustering replicated advance-fee and phishing websites.
\newblock In {\em 2020 IEEE International Conference on Blockchain and Cryptocurrency (ICBC)}, pages 1--8. IEEE, 2020.

\bibitem{ferreira19}
Christof~Ferreira Torres, Mathis Steichen, and Radu State.
\newblock The art of the scam: Demystifying honeypots in ethereum smart contracts.
\newblock In {\em 28th USENIX Security Symposium (USENIX Security 19)}, pages 1591--1607, Santa Clara, CA, August 2019. USENIX Association.

\bibitem{vasek2019analyzing}
Marie Vasek and Tyler Moore.
\newblock Analyzing the bitcoin ponzi scheme ecosystem.
\newblock In Aviv Zohar, Ittay Eyal, Vanessa Teague, Jeremy Clark, Andrea Bracciali, Federico Pintore, and Massimiliano Sala, editors, {\em Financial Cryptography and Data Security}, pages 101--112, Berlin, Heidelberg, 2019. Springer Berlin Heidelberg.

\bibitem{wu2020phishers}
Jiajing Wu, Qi~Yuan, Dan Lin, Wei You, Weili Chen, Chuan Chen, and Zibin Zheng.
\newblock Who are the phishers? phishing scam detection on ethereum via network embedding.
\newblock {\em IEEE Transactions on Systems, Man, and Cybernetics: Systems}, 52(2):1156--1166, 2020.

\bibitem{xia2020characterizing}
Pengcheng Xia, Haoyu Wang, Bowen Zhang, Ru~Ji, Bingyu Gao, Lei Wu, Xiapu Luo, and Guoai Xu.
\newblock Characterizing cryptocurrency exchange scams.
\newblock {\em Computers \& Security}, 98:101993, 2020.

\bibitem{zetzsche2019ico}
Dirk~A Zetzsche, Ross~P Buckley, Douglas~W Arner, and Linus Fohr.
\newblock The {ICO} gold rush: It's a scam, it's a bubble, it's a super challenge for regulators.
\newblock {\em Harv. Int'l LJ}, 60:267, 2019.

\end{thebibliography}

\end{document}